\documentclass[letterpaper,english,10pt, conference, compsocconf]{IEEEtran}
\usepackage[T1]{fontenc}
\usepackage[latin9]{inputenc}
\usepackage{babel}
\usepackage[unicode=true,
 bookmarks=false,
 breaklinks=false,pdfborder={0 0 0},backref=false,colorlinks=false]
 {hyperref}
\hypersetup{pdftitle={User-Constraint and Self-Adaptive Fault Tolerancefor Event Stream Processing Systems},
 pdfauthor={André Martin, Tiaraju Smaneoto, Tobias Dietze, Andrey Brito, Christof Fetzer},
 pdfkeywords={fault tolerance, active replication, passive replication,active standby, passive standby, adaptation, deterministicexecution, precise recovery, gap recovery},
 pdfview=fit,bookmarksdepth=2}

\makeatletter

\pdfpageheight\paperheight
\pdfpagewidth\paperwidth

\usepackage{color}
\usepackage{xcolor,colortbl}
\usepackage{listings}
\usepackage{multirow}
\usepackage{xspace}
\usepackage{tikz}
\usepackage{mathtools}
\usepackage{fixltx2e}
\usepackage{caption}
\usepackage{bbding}
\usepackage{balance}

\newcommand{\fixSpacing}{~\newline\indent}

\renewcommand{\thetable}{\arabic{table}}

\newif\ifcomments 
\commentsfalse

\newif\ifanonym 
\anonymfalse

\ifcomments     
\newcommand{\tentative}[1]{\textbf{\color{blue}#1}}     
\newcommand{\comment}[1]{\textbf{\color{red}#1}} 
\else     
\newcommand{\tentative}[1]{}     
\newcommand{\comment}[1]{} 
\fi

\ifanonym     
\newcommand{\SM}{AnonymESP\xspace} 
\else     
\newcommand{\SM}{\textsc{StreamMine3G}\xspace} 
\fi

\definecolor{methodcolor}{rgb}{0.0,0.1,0.2}
\DeclareTextFontCommand{\mytexttt}{\ttfamily\hyphenchar\font=45\relax}

\makeatother

\begin{document}
\author{

\IEEEauthorblockN{André Martin\IEEEauthorrefmark{1}, Andrey Brito\IEEEauthorrefmark{2},
Christof Fetzer\IEEEauthorrefmark{1}}\\

\IEEEauthorblockA{\IEEEauthorrefmark{1}Technische Universität Dresden,
Dresden, Germany - Email: andre.martin@tu-dresden.de / christof.fetzer@tu-dresden.de}

\IEEEauthorblockA{\IEEEauthorrefmark{2}Universidade Federal de Campina
Grande, Campina Grande, Brazil - Email: andrey@computacao.ufcg.edu.br}\

\vspace{-2ex}}

\title{Elastic and Secure Energy Forecasting\\
in Cloud Environments\vspace{-0.5cm}
}

\maketitle
\begin{abstract}
Although cloud computing offers many advantages with regards to adaption
of resources, we witness either a strong resistance or a very slow
adoption to those new offerings. One reason for the resistance is
that $(i)$~many technologies such as stream processing systems still
lack of appropriate mechanisms for elasticity in order to fully harness
the power of the cloud, and $(ii)$~do not provide mechanisms for
secure processing of privacy sensitive data such as when analyzing
energy consumption data provided through smart plugs in the context
of smart grids. 

In this white paper, we present our vision and approach for elastic
and secure processing of streaming data. Our approach is based on
\SM, an elastic event stream processing system and Intel's SGX technology
that provides secure processing using enclaves. We highlight the key
aspects of our approach and research challenges when using Intel's
SGX technology.\\
\end{abstract}

\maketitle


\section{Introduction \& Background}

\fixSpacing{}As natural resources have become scarce and more costly
throughout the past years, we recently witness a shift from traditional
energy sources to renewable ones such as electricity. Especially with
the advent of new types of energy consumers such as electrical cars,
there is a strong demand for efficient energy management such as carried
out through smart grids.

In smart grids, energy providers try to efficiently distribute energy
by matching their production with the demand by analyzing energy consumption
data provided through smart meters. Such data can be either used to
$(i)$ derive forecasts for future energy consumption or $(ii)$ to
detect anomalies in the grid such as when a smart plug consumes unusual
excessive energy over a longer period of time~\cite{Martin2014b}.

However, providing forecasts and anomaly detection imposes several
challenges: First, the collected data needs to be processed in near
real time in order to react to changing situations in the grid in
a timely manner. Second, the system must be highly scalable as the
data and grid are constantly growing with new smart meters installed
at households each day. Hence, the data processing technology as well
as the infrastructure must be able to grow and scale elastically in
time.

Although scalable real time data processing technologies exit already
for quite some time such as Apache S4~\cite{Neumeyer2010}, Storm~\cite{Storm2014}
and Samza~\cite{Samza2014}, none of them is \emph{elastic} in the
sense that new nodes can be added during runtime where \emph{stateful
components} are transparently moved to those newly acquired resources
without having to either tear down or restart the system.

On infrastructure side, cloud computing is a new paradigm that addresses
the scalability requirement for such dynamic applications as resources
can be conveniently added or released at any point of time. However,
cloud computing raises several \emph{privacy} issues as the data is
being processed in a shared environment in data centers which are
not under the administrative control of the energy provider driving
the analysis. Although there exist already mechanisms to securely
transfer and store data in cloud environments, processing data in
cloud environments is still an open issue. 

In this white paper, we present our vision for \emph{elastic} and
\emph{secure} real time data processing in cloud environments using
\SM~\cite{StreamMine3G2016}, an elastic data stream processing
engine paired with Intel's latest SGX technology~\cite{SGX2015}.\\

\section{Elastic Data Processing}

\fixSpacing{}During the past years, an increasing number of businesses
is moving from traditional batch processing to online processing using
data streaming systems. However, processing data in a streaming fashion
requires the use of state which imposes new challenges when the system
needs to be elastic. Elasticity requires that the computation and
its associated state can be moved (i.e., \emph{migrated}) to a new
machine during the course of processing. However, moving state to
a new node in a consistent manner is not trivial: First, the stream
of data tuples (i.e., events) must be redirected to the new location,
i.e., the spare node. Second, the state must be transferred to the
new location, however, in such a way that neither events are lost
nor processed twice due to the redirected data stream.

Unfortunately, the majority of nowadays MapReduce~\cite{Dean2008}
inspired stream processing systems lack support for elasticity completely.
Either state is not managed at all such as in Apache Storm~\cite{Storm2014}
or the systems lack appropriate operator and state migration mechanisms
(Apache S4~\cite{Neumeyer2010}).

In this white paper, we advocate for our \SM~\cite{StreamMine3G2016}
approach where we utilize a combination of mechanisms such as \emph{deterministic
processing \& merge} and \emph{active replication} in order to achieve
elasticity with strong consistency guarantees, i.e., exactly once
processing semantics. The key idea of our approach is as follows:
For an operator migration, we create a second replica with a virgin
state first. In addition to the operator replica, the incoming event
stream is duplicated, i.e., both replicas receive identical data now.
Next, a snaphot-copy of the state from the original one is taken and
transferred to the second replica. The state has a timestamp vector
associated which allows the system to identify obsolete events (duplicates).
Once the two replicas are in sync, i.e., produce identical sets of
events, the original operator instance is torn down leaving only the
second replica in the system.

In order to ensure consistency across replicas, we enforce a strict
ordering of incoming events. This requires that the incoming events
are associated with strong monotonically increasing timestamps. Those
timestamps are then used to established a total order using a priority
queue at each of the replicas.

The previously described protocol allows to migrate operators without
any interruption of the event stream as well as noticeable latency
increases. However, requires ordering of events which imposes a non-negligible
overhead. Note that the overhead of ordering can be reduced by exploiting
operator properties such as commutativity as we have shown in previous
work~\cite{Martin2011}.\\

\section{Secure Data Processing}

\fixSpacing{}Although cloud computing is an appealing technology
as it reduces infrastructure costs for highly dynamic applications
such as stream processing, there is only a slow adoption of it as
it imposes high risks with regards to privacy and confidentiality.
Especially when processing private data such as energy consumption
records collected through smart meters, cloud computing cannot ensure
the required privacy for such types of applications. However, a technology
that closes this gap is Intel SGX where certain parts of the program
code can be run in a so called \emph{enclave}.

An enclave is a \emph{trusted environment }which is not accessible
by any program code outside of the enclave such as potential malicious
third party applications. Furthermore, remote attestation ensures
that the code loaded and running inside the enclave has not been tampered
with.

For a secure data processing in cloud environments, we envision to
utilize Intel SGX in \SM. The key idea of our approach is to run
solely the operator code inside the enclave in a light-weight manner
rather than the complete software stack of the data streaming system.
Data exchanged between operator instances or with entities outside
of \SM is encrypted prior leaving the trusted environment.

However, the use of such enclaves imposes several challenges: First,
the so called enclave page cache (EPC) is limited to $128MB$ in size,
hence, applications that utilize larger portions of state such as
when providing an energy consumption forecast based on historical
data need efficient swapping/cache eviction mechanisms which must
be thoroughly evaluated in the context of stream processing applications.
Second, operators running in enclaves must be also sufficiently protected
against potential attacks through the system call interface. For the
latter one, we envision the use of a modified version of libmusl~\cite{LibMusl2016},
equipped with an additional protection layer where function arguments
and return values are checked prior passing them between the trusted
and untrusted environment.

Since Intel SGX is available on all newly distributed Intel CPUs (Skylake
architecture) since fall 2015, the approach does not require the acquisition
of special and expensive hardware to achieve secure data processing
in untrusted environments such as when using secure co-processors.\\

\section{Conclusion \& Summary}

\fixSpacing{}In this white paper, we presented an approach for elastic
and secure data processing in cloud environments targeting privacy
sensitive applications such as used when processing smart meter data
originating from smart grids. Our approach is based on \SM~\cite{StreamMine3G2016},
a data streaming system that provides elasticity for stateful operators
through a combination of active replication and deterministic processing.
In order to ensure privacy, we execute operators in \SM inside of
Intel SGX enclaves and enforce encryption of data for all outgoing
and incoming data tuples. We believe that this approach is a promising
direction as it will allow us to run privacy sensitive big data applications
in untrusted cloud computing environments.

\section*{Acknowledgment}

\fixSpacing{}\fontsize{8}{9}\selectfont{The research leading to
these results has received funding from the European Community's Framework
Program Horizon 2020 under grant agreement number 690111 (SecureCloud),
692178 (EBSIS) and by the German Excellence Initiative, Center for
Advancing Electronics Dresden (cfAED), Resilience Path.}

\balance\setlength{\parskip}{1.0\baselineskip}

\section*{Team Profile}

\emph{Christof Fetzer} is at professor at TU Dresden, Germany where
he is leading the Systems Engineering group since 2004. He has more
than 25 years of experience in distributed systems and dependability
research. \emph{Andrey Brito} is a professor at Universidade Federal
de Campina Grande in Brazil. His research focus is cloud computing,
stream processing systems with an emphasis on scalability and dependability.
\emph{André Martin} finished his PhD (December 2015) at TU Dresden.
In his PhD thesis, he explored novel mechanisms for dependability
and elasticity in streaming systems. He is also the creator and maintainor
of \SM. The team participated twice succesfully in the DEBS challenge
(2014 \& 2015) and also won the \emph{UCC2014 Cloud Challenge award}~\cite{Martin2014b}
where they were showcasing the elasticity mechanisms for stream processing.

\bibliographystyle{IEEEtran}
\bibliography{Paper}

\end{document}